\documentclass[twocolumn]{revtex4}

\usepackage{graphicx}
\usepackage{color}
\usepackage{natbib}
\usepackage{amssymb}
\usepackage{amsfonts}
\usepackage{amscd}
\usepackage{latexsym}
\usepackage{amsmath}
\usepackage{amsbsy}
\usepackage{amsgen}
\usepackage{pifont}
\usepackage[english]{babel}
\usepackage[utf8]{inputenc}
\graphicspath{{./pics/}}

{\rule{20pt}{1ex}\endlist}
\let\oldmarginpar\marginpar
\renewcommand\marginpar[1]{\-\oldmarginpar[\raggedleft\footnotesize #1]%
{\raggedright\footnotesize #1}}

\makeatletter                                  
\def\DonormalEndcol{%
\ifx\toporbotfloat\xtopfloat%
  \ifcaptypefig%
  \expandafter\gdef\csname topfloat\the\figandtabnumber\endcsname{%
  \vbox{\vskip\PushOneColTopFig%
  \unvbox\csname figandtabbox\the\loopnum\endcsname%
  \vskip\abovefigcaptionskip%
  \csname caption\the\loopnum\endcsname%
  \csname letteredcaption\the\loopnum\endcsname%
  \csname continuedcaption\the\loopnum\endcsname%
  \csname letteredcontcaption\the\loopnum\endcsname            
  \ifredefining%
  \csname label\the\loopnum\endcsname%
  \expandafter\gdef\csname topfloat\the\loopnum\endcsname{}\fi}%
  \vskip\intextfloatskip
  \vskip-4pt 
}%
\else%
  \ifcaptypeplate%
  \expandafter\gdef\csname topfloat\the\figandtabnumber\endcsname{%
  \vbox{\vskip\PushOneColTopFig%
  \unvbox\csname figandtabbox\the\loopnum\endcsname            
  \vskip\abovefigcaptionskip                           
  \csname caption\the\loopnum\endcsname                    
  \csname letteredcaption\the\loopnum\endcsname                
  \csname continuedcaption\the\loopnum\endcsname               
  \csname letteredcontcaption\the\loopnum\endcsname            
  \ifredefining                                
  \csname label\the\loopnum\endcsname                      
  \expandafter\gdef\csname topfloat\the\loopnum\endcsname{}\fi}        
  \vskip\intextfloatskip 
  \vskip-4pt 
}%
\else
 \expandafter\gdef\csname topfloat\the\figandtabnumber\endcsname{%
 \vbox{\vskip\PushOneColTopTab 
 \csname caption\the\loopnum\endcsname                     
  \csname letteredcaption\the\loopnum\endcsname                
  \csname continuedcaption\the\loopnum\endcsname               
  \csname letteredcontcaption\the\loopnum\endcsname            
  \vskip\captionskip                               
  \unvbox\csname figandtabbox\the\loopnum\endcsname            
\ifredefining                                  
\csname label\the\loopnum\endcsname                    
\expandafter\gdef\csname topfloat\the\loopnum\endcsname{}\fi           
}\vskip\intextfloatskip 
\vskip-10pt}                                   
\fi\fi%
\else
\ifcaptypefig                                  
\expandafter\gdef\csname botfloat\the\figandtabnumber\endcsname{%
\vskip\intextfloatskip                             
\vbox{\unvbox\csname figandtabbox\the\loopnum\endcsname            
\vskip\abovefigcaptionskip                         
  \csname caption\the\loopnum\endcsname                    
  \csname letteredcaption\the\loopnum\endcsname%
  \csname continuedcaption\the\loopnum\endcsname%
  \csname letteredcontcaption\the\loopnum\endcsname%
\vskip\PushOneColBotFig
\ifredefining%
\csname label\the\loopnum\endcsname                    
\expandafter\gdef\csname botfloat\the\loopnum\endcsname{}\fi}}%
\else                                      
\ifcaptypeplate                                
\expandafter\gdef\csname botfloat\the\figandtabnumber\endcsname{%
\vskip\intextfloatskip                             
\vbox{\unvbox\csname figandtabbox\the\loopnum\endcsname            
\vskip\abovefigcaptionskip                         
  \csname caption\the\loopnum\endcsname                    
  \csname letteredcaption\the\loopnum\endcsname%
  \csname continuedcaption\the\loopnum\endcsname%
  \csname letteredcontcaption\the\loopnum\endcsname%
\vskip\PushOneColBotFig
\ifredefining%
\csname label\the\loopnum\endcsname                    
\expandafter\gdef\csname botfloat\the\loopnum\endcsname{}\fi}}%
  \else
\expandafter\gdef\csname botfloat\the\figandtabnumber\endcsname{%
  \vskip\intextfloatskip                           
\vbox{\csname caption\the\loopnum\endcsname                
  \csname letteredcaption\the\loopnum\endcsname                
  \csname continuedcaption\the\loopnum\endcsname               
  \csname letteredcontcaption\the\loopnum\endcsname%
  \vskip.5\intextfloatskip                         
  \unvbox\csname figandtabbox\the\loopnum\endcsname%
\vskip\PushOneColBotTab                            
\ifredefining%
\csname label\the\loopnum\endcsname                    
\expandafter\gdef\csname botfloat\the\loopnum\endcsname{}\fi}}%
\fi\fi\fi}                                 
\makeatother                                   

\begin{document}
\title{In-drop capillary spooling of spider capture thread inspires highly extensible fibres}
\author{Herv\'e Elettro}
\email{elettro@dalembert.upmc.fr}

\author{S\'ebastien Neukirch}

\author{Arnaud Antkowiak}

\affiliation{Sorbonne Universités, UPMC Univ Paris 06, CNRS, UMR 7190 Institut Jean Le Rond d'Alembert, F-75005 Paris, France.}

\author{Fritz Vollrath}
\affiliation{Oxford Silk Group, Zoology Department, University of Oxford, UK}
\date{\today}

\begin{abstract}
Spiders’ webs and gossamer threads are often paraded as paradigms for lightweight structures and outstanding polymers. Probably the most intriguing of all spider silks is the araneid capture thread, covered with tiny glycoprotein glue droplets. Even if compressed, this thread remains surprisingly taut -- a property shared with pure liquid films --  allowing both thread and web to be in a constant state of tension. Vollrath and Edmonds proposed that the glue droplets would act as small windlasses and be responsible for the tension, but other explanations have also been suggested, involving for example the macromolecular properties of the flagelliform silk core filaments. Here we show that the nanolitre glue droplets of the capture thread indeed induce buckling and coiling of the core filaments: microscopic in-vivo observations reveal that the slack fibre is spooled into and within the droplets. We model windlass activation as a structural phase transition, and show that fibre spooling essentially results from the interplay between elasticity and capillarity. This is demonstrated by reproducing artificially the mechanism on a synthetic polyurethane thread/silicone oil droplet system. Fibre size is the key in natural and artificial setups which both require micrometer-sized fibres to function. The spools and coils inside the drops are further shown to directly affect the mechanical response of the thread, evidencing the central role played by geometry in spider silk mechanics. Beside shedding light on araneid capture thread functionality, we argue that the properties of this biological system provide novel insights for bioinspired synthetic actuators.
\end{abstract}

 \maketitle
    

The orb-webs of araneid spiders are highly effective and efficient traps that are able to fully cushion the impact of an insect within a second and then manage to keep the prey well ensnared in a wet-sticky capture spiral until the spider can reach it \citep{Foelix2010}. 
%
%
The exceptional extensional properties of the spiral threads rely on the macromolecular architecture of the silk fibroins \citep{Becker2003,Blackledge2005} while the ability to contract, self-tense and adapt slack length with virtually no hysteresis has been attributed to a surprisingly complex mechanical micro-mechanism \citep{Vollrath1989}.   
%
Here we are setting out to test this micro-windlass hypothesis. To date it is based mostly on indirect snap-shot observations of various amounts of coiled threads inside capture spiral droplets laid down on a microscope slide \citep{Vollrath1989} combined with the interpretation of mechanical behaviour. 
The lack of direct, dynamic observation of the phenomenon has given rise to criticism refuting the conclusions of an active mechanism driving the system  (see  \citep{Schneider1995} with reply \citep{Vollrath1995} and summary \citep{Peters1995}).  Much of the criticism was dismissed by empirical studies \citep{Vollrath1995} but a fundamental question must be taken seriously: after all, the physics of such a mechanism would require the watery coating of the thread to be able to \textit{(i)} exert forces sufficiently powerful to rapidly tension substantial lengths of the filament which bears hundreds of nanolitre droplets, and \textit{(ii)} hold the composite system in place without sagging against gravity.  
%
%
Actually, the vast literature on liquid coating on fibres, whether in textile \citep{Adam1937} or glass filaments \citep{Quere1999} products or indeed in biological applications such as the wetting of mammalian hairs  \citep{Carroll1989}
, does not contain any reports of such drop-induced 'windlass' events. 

Our results fully support the windlass theory and identify the fundamental mechanism to be dependent on the interplay between elasticity and capillarity.  Moreover, we demonstrate empirically as well as theoretically that this mechanism is generic, \textit{i.e.} not requiring silk proteins in either the filaments nor the droplets as fundamental material components.  This not only elucidates the physics of the system but also opens the way for the design of novel bio-inspired synthetic actuators using the windlass array concept. 

\begin{figure*}[ht]
\begin{center}
\includegraphics[width=183mm]{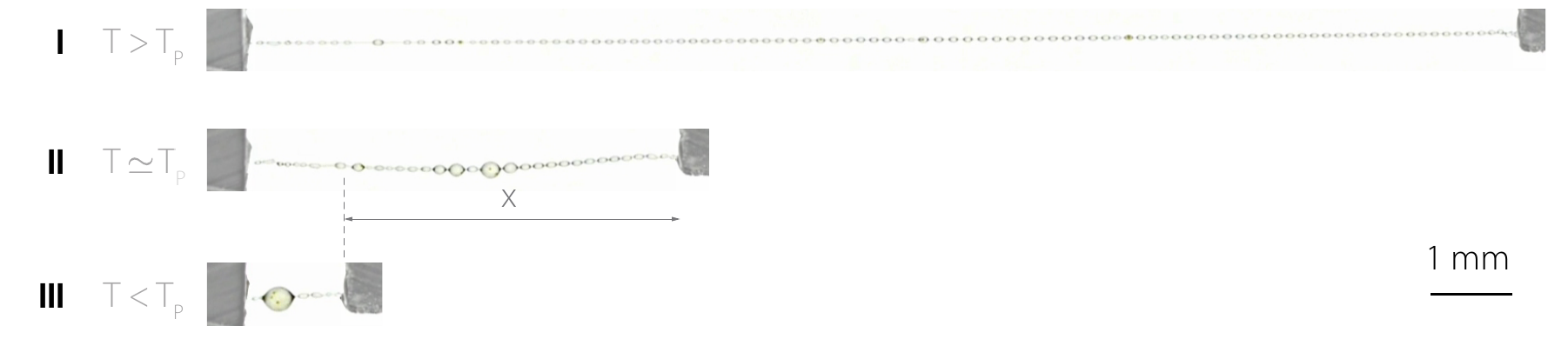}
\caption{\textbf{A liquid-like fibre}. Whether stretched or relaxed, the typical capture silk thread of an araneid orb spider (here \textit{Nephila edulis}) remains taut.
Force monitoring reveals a threshold tension $T_\text P$ above which the fibre behaves like a spring (\textsf{I}).
As the force is decreased a force plateau $T \simeq T_\text P$ is reached, along which the fibre adopts a wide range of lengths, just as soap films do (\textsf{II}).
At lower forces, $T < T_\text{P}$, the fibre is totally contracted (\textsf{III}).  See also Supplementary Videos for full cycle.}
\label{fig:natural-caliper}
\end{center}
\end{figure*}

Figure~\ref{fig:natural-caliper} outlines the macroscopic observation of the system.  The natural set-up is fully self-assembling, with the spider coating the core fibres of the capture spiral with a thin layer of hygroscopic compounds \citep{Vollrath1991}. This layer then swells by taking up water from the atmosphere and go through a Plateau-Rayleigh instability to form a string of interconnected silk droplets sitting astride the core silk fibres \citep{Edmonds1992}. This natural micron-sized windlass array system ensures that the capture thread is always under tension -- independent of any extension or relaxation.  As the capture thread is stretched, lengths of core filament are simply strained or, depending on past action, pulled out of a droplet.  It appears that the surface tension of the droplet is strong enough to pull-in the silk filament, which buckles and then coils up inside the liquid.

\begin{figure}[t]
\begin{center}
\includegraphics[width=89mm]{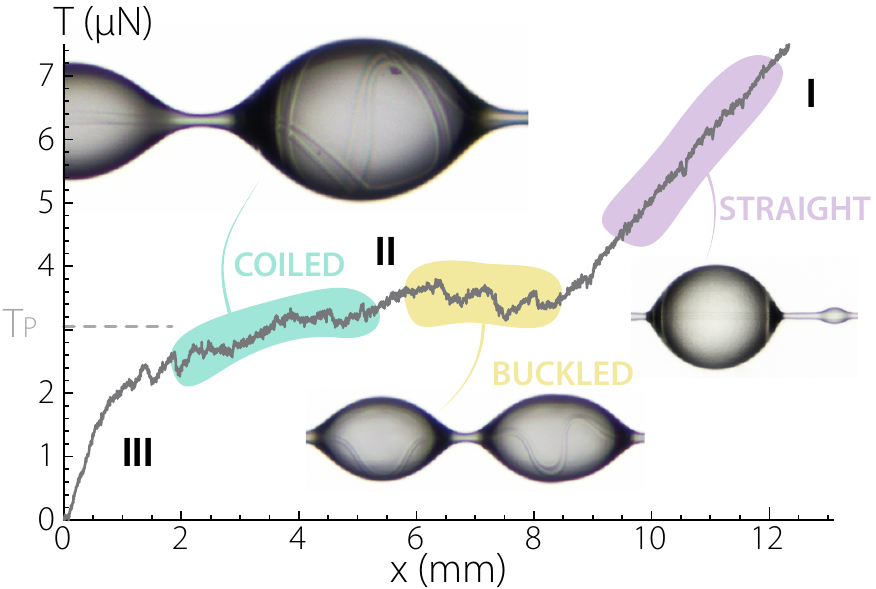}
\caption{\textbf{Shape-induced functionalization.} Quasi-static force measurements on spider capture threads combined with microscopic observations demonstrate that the slack core filament is spooled into the droplets (typically 250-300 $\mu$m wide) along the force plateau $T = T_\text P$. 
%
For larger forces $T>T_\text P$, the fibre straightens and a spring-like behaviour is shown.
The characteristic J-shape for this force-extension curve can be attributed to a shape-induced functionalization of the fibre by the glue droplets (see also Supplementary Video).}
\label{fig:natural-jcurve}
\end{center}
\end{figure}

\begin{figure}[t]
\begin{center}
\includegraphics[width=89mm]{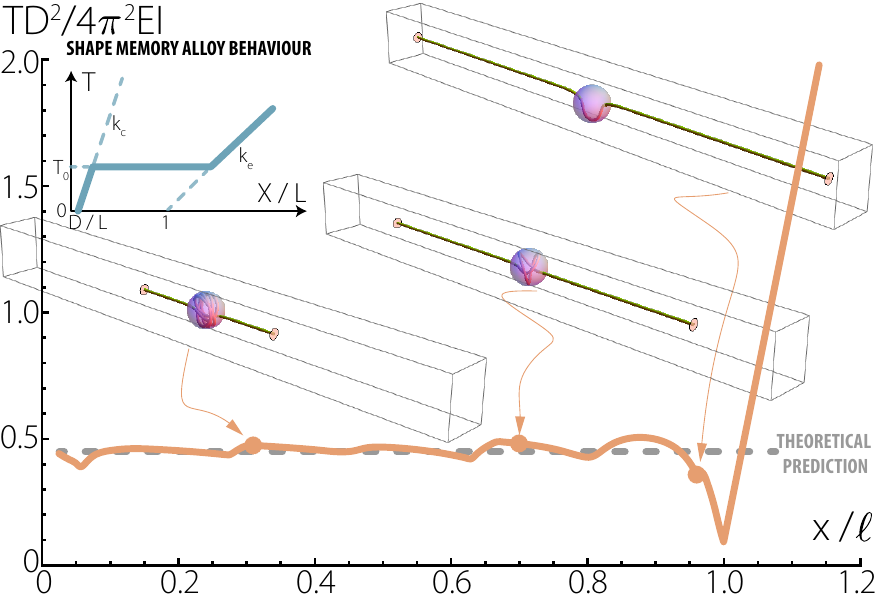}
\caption{\textbf{Windlass as a structural phase transition.} Numerical simulations of a slender fibre supporting a drop sitting astride unravel a subcritical buckling transition where the slack fibre is suddenly localised within the droplet.
Continuation of equilibria, with $F_\gamma D^2 / (4\pi^2 \, EI) = 0.5$ and $L=20\, D$, reveal the fine details in the micro-mechanical response, resulting in inhomogeneities in the Maxwell plateau.
The approximate plateau value $T_\text{P} D^2 / (4\pi^2 \, EI) \simeq 0.45$, given by equation~(\ref{equa:T-analytic}), is drawn for comparison.
Several equilibrium configurations illustrate the coiling of the fiber and its strong packing within the droplet.
The mechanical response is  typical of systems exhibiting first-order phase transitions (such as shape-memory alloys - SMA), and the numerical results are confirmed by an analytical model involving a wet/coiled (analog to the austenite phase in SMA) and a dry/stretched (analog to the martensite phase in SMA) phase (see inset).}
\label{fig:sma}
\end{center}
\end{figure}

Figure~\ref{fig:natural-jcurve} shows force-elongation measurements performed on a single capture thread, accompanied by microscopic observations of core fibre coiling within the glue droplets. This data  reveals the self-organisation of the core filaments into coils within the droplet, and determines the link between the shape of the filament and the mechanical response of the thread. 
Indeed the observed mechanical response encompasses three different stages, each characterized by a typical force-displacement relation and a specific filament shape within the drop. At very low tension $T$ exerted on the fibre, the central glue droplet, resulting from the successive merging of tiny drops, has `ingested' a large amount of spare fibre and presents a roughly linear relation between force and extension, see region \textsf{III}. As a plateau force $T\simeq T_\text P$ is reached, the filament starts uncoiling and exiting the drop, allowing this composite system to effortlessly reach very large extensions, see region \textsf{II}. Interestingly this mechanical response, associated with unspooling, bears little resemblance with the typical response of a fibre, and is much more reminiscent of the mechanics of soap films  \citep{Gennes2003}. Past the threshold where all the coils have been straightened, the capture thread recovers a classic drop-on-(straight) fibre appearance along with a linear stretched spring regime response, see region \textsf{I}. 
This trimodal response of the capture thread is not unlike the typical response of a Shape Memory Alloy (SMA) fibre, see Figure~\ref{fig:sma}. And indeed in both systems the plateau regime corresponds to a phase transition: from austenite to martensite in SMA, and from coils-in-drop to straight fibre in capture silk.

\noindent To elucidate the conditions for the windlass action, we examine this phase transition further. We start by considering a model capture thread composed of a soft fibre capable of being bent or stretched and subjected to an external tension $T$, supporting a wetting liquid drop standing astride \citep{Lorenceau2006}. Though thin, silk filaments are not expected to be significantly excited with thermal fluctuations: the persistence length of such filaments is indeed estimated to be kilometer-sized  \citep{Marko2003}, to be compared with a few tens of nanometers for DNA. Disregarding entropic elasticity effects, we describe windlass activation as a phase transition between a coiled phase -- where the fibre is entirely packed within the liquid drop -- and an extended phase -- where the thread runs straight outside the drop.
The extended phase is modelled as a spring with stiffness $k_\text e = \pi E h^2 /\ell$, where $\ell$ stands for the rest length, $E$ the     Young's modulus of the fibre and $h$ its radius. The strain energy of the extended phase is then $\frac{1}{2} k_\text e (x_\text e - \ell)^2$, where $x_\text e$ is the phase's extension.
The coiled phase typically describes the system at low applied tension~$T$. In the limit $T=0$ the coiled phase extension is $x_\text c=D$ where $D$ is the diameter of the, then spherical, liquid drop. As tension starts rising, the liquid drop deforms into an elongated shape. Modelling the deformation as ellipsoidal yields the following spring-like relation: $T=\frac{4}{5} \pi \gamma (x_\text c - D)$, where $\gamma$ is the liquid surface tension. The strain energy of the coiled phase is then ${\cal E}_\text c = \frac{1}{2} k_\text c (x_\text c - D)^2$ where the spring stiffness of the liquid drop directly arises from surface tension $k_\text c=\frac{4}{5} \pi \gamma$.
As typical for phase transition problems, there is an energetical cost per unit length associated with the transformation from the coiled phase to the extended phase, which we note $\epsilon_0$. This energy originates from both the uncoiling of the fibre and the energy difference between wet and dry states. The uncoiling process results in an elastic energy gain $-\frac{1}{2} EI \kappa^2$ per unit length, where the curvature $\kappa = 2/D$ and $I= \frac{\pi}{4} h^4$. The latent energy per unit length also embodies the difference in surface energies: in the coiled phase the fibre is surrounded by liquid, hence bears a surface energy $2 \pi h \gamma_{\text{sl}}$ per unit length whereas in the extended phase the air surrounding the fibre yields an energy $2 \pi h \gamma_\text{sv}$ per unit length. Here $\gamma_{\text{sl}}$ and $\gamma_\text{sv}$ denote solid-liquid and solid-air interface energies respectively. It is to be noted that a surface energy constrast between the coiled and extended phase persists even if a thin liquid film sheathes the core filament, for the energy of a filament coated with a thin liquid layer differs from that of a filament immersed in liquid  \citep{Gennes2003}. Upon using Young-Dupr\'e wetting relation $\gamma_\text{sv}-\gamma_{\text{sl}} = \gamma \cos \theta$, the energetical cost for phase transformation is obtained: $\epsilon_0 =2 \pi h \gamma \cos \theta - \frac{1}{2}  \pi E  h^4/D^2$, where $\theta$ is the liquid contact angle on the fibre. From this expression we readily obtain conditions for windlass activation. Indeed, for the coiled phase to be stable at small extensions $x$, $\epsilon_0$ has to be positive. This condition can be recast into a condition for the radius, where it becomes apparent that only extremely thin fibres can deploy a windlass:
\begin{equation}
h < \left( \frac{4 \gamma \cos \theta}{E} \right)^{1/3}  D^{2/3}
\end{equation}
Noting the total energy ${\cal E}_e$ of the extended phase as ${\cal E}_\text e~=~\frac{1}{2}k_\text e~(x_\text e-\ell)^2+\epsilon_0\ell$, we write the global energy when part of the fibre is in the coiled phase, and part in the extended phase.  Denoting the extended phase fraction $\rho$, the global energy ${\cal E}(x_\text e,x_\text c,\rho)$ reads:
${\cal E}~=~(1-\rho){\cal E}_\text c(x_\text c)+\rho{\cal E}_\text e(x_\text e)$.
%
Minimizing this energy we find that the fibre can be entirely in the coiled phase ($\rho = 0$; filament fully packed in the drop) with tension $T = k_\text c (x - D)$, or entirely in the extended phase ($\rho=1$) with tension $T = k_\text e (x - \ell)$. A third interesting possibility consists in a mixture of phases $0<\rho<1$. In this latter case, part of the filament is packed in the drop but the outer part is taut, consistent with our observations. In the limit where $D \ll \ell$ and $\epsilon_0 \ll k_\text e, k_\text c$, we have $x_\text c \simeq D+\epsilon_0/k_\text c$, $x_\text e \simeq \ell + \epsilon_0/k_\text e$ and $T \simeq T_\text{P} \simeq \epsilon_0$, so that the plateau tension is
\begin{equation}
T_\text{P} \simeq 2 \pi h \gamma \cos \theta - \frac{1}{2} \pi E \frac{h^4}{D^2} \label{equa:T-analytic}
\end{equation}
We performed detailed numerical computations of equilibrium of an extensible and flexible elastic fibre \citep{Audoly2010,Antkowiak2011}. The filament, held at both extremities with imposed distance $x$, was subjected to attracting meniscus forces $F_{\gamma}$ at entrance and exit of a confining sphere. The loading $(x,T)$ diagram, shown in Fig.~\ref{fig:sma}, reveals inhomogeneities in the Maxwell line. These inhomogeneities are due to fine details in the micro-mechanical response (spooling and unspooling events) of the system. 
Setting $F_\gamma = 2 \pi h \gamma \cos \theta$, we plot in Fig.~\ref{fig:sma} the horizontal line corresponding to equation~(\ref{equa:T-analytic}), and see that it compares well with numerical computations.

\begin{figure}[t]
\begin{center}
\includegraphics[width=89mm]{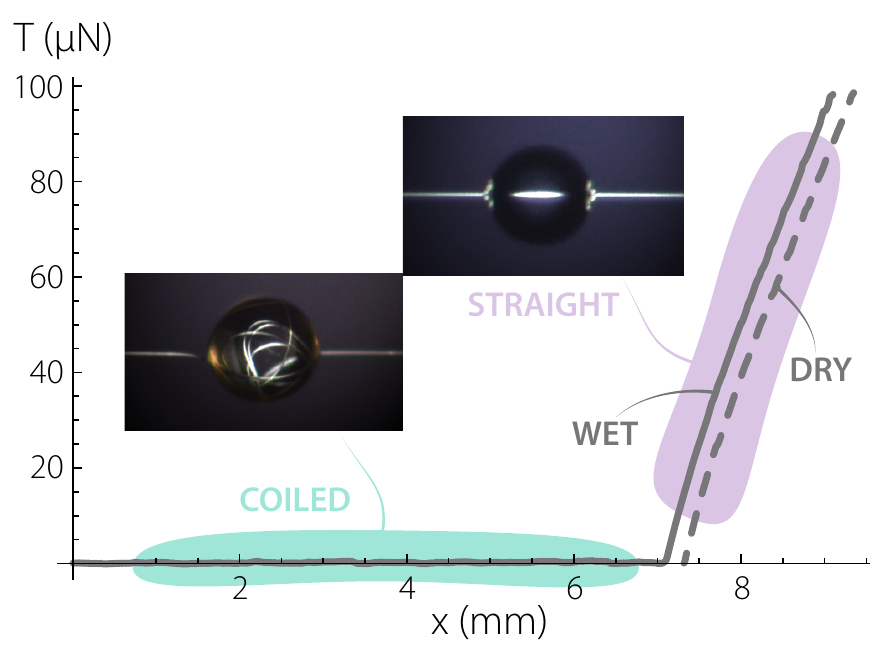}
\caption{\textbf{A synthetic windlass mechanism.} Combined quasi-static mechanical measurements and polarizer-analyzer observations of silicone oil droplet sitting astride thin TPU fibre ($h=2.3\pm 0.15 \mu$m, $E=17 \pm 3$MPa, and wetting angle $\theta=36\pm 7^\circ$ for the left inset) demonstrate that the key features of the spider capture thread can be obtained with an artificial system. The natural mechanical response of TPU, shown in dashed line, is significantly altered by the addition of an oil droplet (wet length $380 \mu$m for the left inset), resulting in a highly extensible system (here $+9000 \%$ breaking strain). The resulting J-shape curve appears to be associated with elastocapillary spooling within the oil droplet, in agreement with our theoretical description (see also Supplementary Video).}
\label{fig:arti-jcurve}
\end{center}
\end{figure}

\begin{figure*}[t]
\begin{center}
\includegraphics[width=183mm]{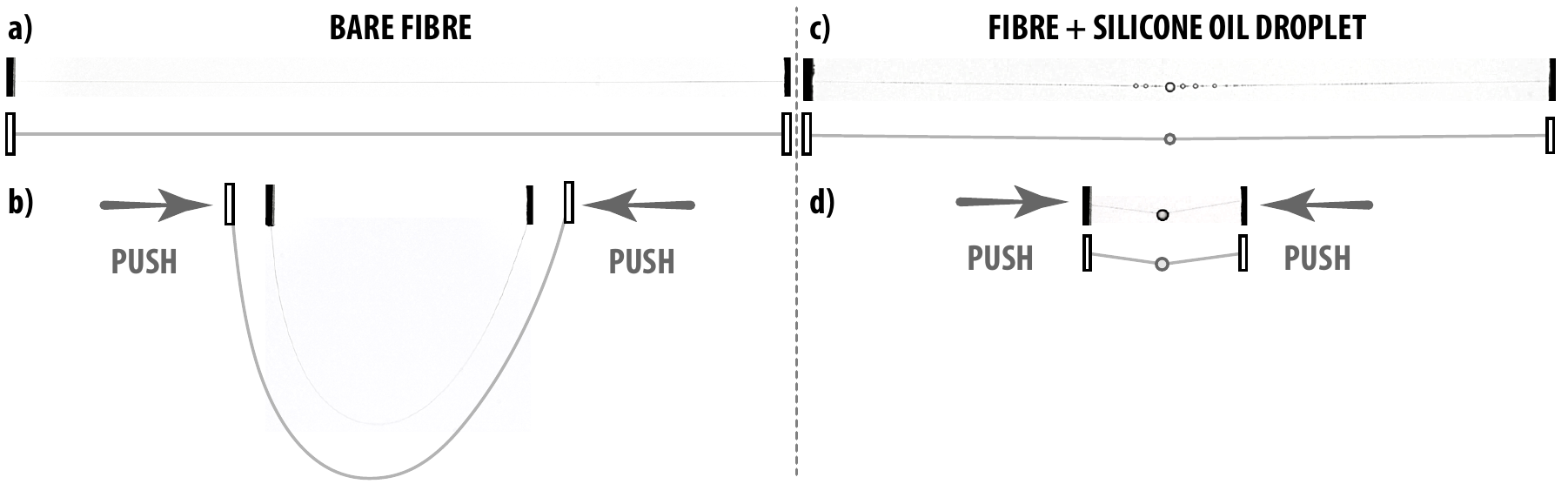}
\caption{\textbf{Artificial spider capture thread.} Left: a thin micrometer-sized TPU filament is held on a caliper. Bringing  the ends together leads the filament to exhibit a characteristic catenary-like sagging shape (note: the outlines serve as guides for the eye). Right: This behaviour changes after deposition of a small silicone oil droplet on the filament: bringing the extremities closer leaves the filament taut while the excess filament enters the central droplet.}
\label{fig:arti-caliper}
\end{center}
\end{figure*}

Our theoretical results show that, provided capillarity prevails over filament elasticity, elastocapillary spooling can be initiated on any flexible fibre, and that there is no need for any rearrangement in the molecular architecture. We experimentally demonstrated this material independence by triggering windlass on a soft Thermoplastic PolyUrethane (TPU) micron-sized filament. As soon as a single silicone oil droplet 
was deposited on the sagging TPU filament, the fibre straightened and a tension force on the fibre of typically  1 $\mu$N immediately developed. We performed tensile tests and observed that the force curve, see Fig.~\ref{fig:arti-jcurve}, exhibited the plateau regime with $T_\text{P} \sim 1 \mu$N until the filament was totally straight and the spring regime started. Snapshots of the filament within the oil droplet in the plateau and spring regimes are shown in Fig.~\ref{fig:arti-jcurve} and confirm the link between filament shape and the mechanical response of the system.
In addition we show in Fig.~\ref{fig:arti-caliper} that a synthetic TPU fibre displayed the same overall properties of hypercontraction than capture silk: the liquid drop put the system in tension and there was virtually no sagging, even up to fivefold length reduction.

Interestingly the addition of the oil droplet allows to turn (and tune) the mechanical response into the J-shaped curve so typical of many collagenous tissues \citep{Gordon1978}. Mimicking the mechanical response of biological tissues with synthetic materials certainly offers appealing perspectives for e.g. the design of artificial muscles.
Actually, more than the material properties, this is the shape of the strongly curled spools of core filament, reminiscent of packed DNA in bacteriophages \citep{Leforestier2009}, that drives the mechanical response of both the natural and the artificial systems. These typical examples of shape-induced functionalization \citep{Ball2009} illustrate the fact that geometry plays a key role in spider silk and TPU/oil droplet 'extreme' mechanics \citep{Lazarus2012,Krieger2012}.

In conclusion we have shown that the mechanical response of capture silk relies on it being a thin fibre decorated with drops powering the micro-windlass action. Thus the extraordinary mechanical behaviour of araneid spider capture silk relies more on the macroscopic composite structure than on the microscopic material properties, underlining that natural selection 
sometimes cleverly tinkers by coopting simple laws of physics \citep{Jacob1977}.

\subsection*{Materials}

\textbf{Capture silk samples.} Our \textit{Nephila edulis} spider was kept in a $80\times80\times30$ cm vivarium, consisting of wood panels, PMMA windows and artificial plants. The spider was kept at high humidity (above 70\%) and comfortable temperature (above $22^\circ$C)  with a 12/12 hr day/night schedule. The spider was fed crickets and flies three times a week. Sections of web were carefully excised  using a soldering iron for transfer within a rigid frame. To visualise the fibre running through each droplet, the humidity was set to 100\% rH for 15 minutes, then stepped down to 50\% before observation.\\
\textbf{Artificial samples.} PolyUrethane (TPU, Elastollan 1185A from BASF\textsuperscript{\textregistered}) granules were deposited on a hot plate at $230^\circ$C. After melting, we used a tweezer to pick up a small droplet which  was then stretched quickly while at the same time being released into ambient room temperature. This resulted in the creation of micron-sized, metre-long, soft filaments. The filament was then deposited on the measuring setup as outlined below. A droplet of silicone oil (Rhodorsil\textsuperscript{\textregistered} 47V1000) was then deposited by gently touching and brushing the filament with a drop hanging from a pipette.\\
\textbf{Measurement methods.} Filament samples were transferred to the measuring setup by coiling one end around the tip of a FemtoTools FT-FS1000 capacitive deflection force sensor (50 nN-1mN range) and gluing the other end to a glass slide as base. The force sensor was mounted on a linear micro positioner SmarAct SLC-1730 (repetability 0.5 $\mu$m) and measurements are performed through a work station by USB connection. All the tests were performed in stretching at a speed of $25 \mu$m/s, and considering the centimeter size in length of the sample, they can be considered to be quasi-static. The optical setup consisted of a Leica macroscope (VZ85RC) mounted on a micro-step motor and a 3 megapixels Leica DFC-295 camera ($400\times$ zoom, 334 nm/pixel picture resolution) or a D800E Nikon camera with 3 10mm C-mount extension rings (937 nm/pixel video resolution and 374 nm/pixel photo resolution) alternatively. We used a Phlox 50x50 mm backlight, at 60000 lux or alternatively an optical fibre with LED lamp (Moritex MHF-M1002) with circular polarizer. Side views were acquired with a second D800E camera, with a 70mm extension tube and a 100mm macro Zeiss lense (7,27 microns/pixel video resolution).
The force sensor was tared to zero with the fibre compressed slightly more than its slack length, so that it sags, but only minutely, be it for fibres with or without droplet. The measurement of the slack length was performed by pulling on the filament at one end by a few micrometers to straighten the fibre.\\
The TPU fibre diameter measurement was performed using Fiji software. A high-resolution picture of the fibre is analysed using the following steps : the contrast is enhanced up to the point that 0.4\% of the pixels are saturated, then the grey value of the pixels on a line perpendicular to the fibre axis is plotted. The typical curve obtained this way resembles a downward pointing gaussian, thus the diameter of the fibre is extracted as the full width at half minimum of the peak.\\
\textbf{Numerical computations.}
The windlass system is modeled as an elastic filament, obeying Kirchhoff equilibrium equations, in interaction with a sphere. Except at the two `meniscus' points, the filament is prevented from touching or crossing the sphere through a soft-wall barrier potential. The equilibrium of the system is solved using two-points boundary-value problem techniques (shooting method in Mathematica, and collocation method using the Fortran - AUTO code).
\subsection*{Supplementary information}

The coiled phase has extension $x_c$. It is modeled as a linear spring with stiffness $k_c=4/5 \, \pi \gamma$, where $\gamma$ is the liquid surface tension. The tension-extension relation is then
\begin{equation}
T=4/5 \, \pi \, \gamma (x_c - D)
\end{equation}
where $D$ is the diameter of the drop at $T=0$. The strain energy of the coiled phase is then
\begin{equation}
{\cal E}_c = 1/2 \, k_c \, (x_c - D)^2
\end{equation}

\begin{figure}[h]
\begin{center}
\includegraphics[width=80mm]{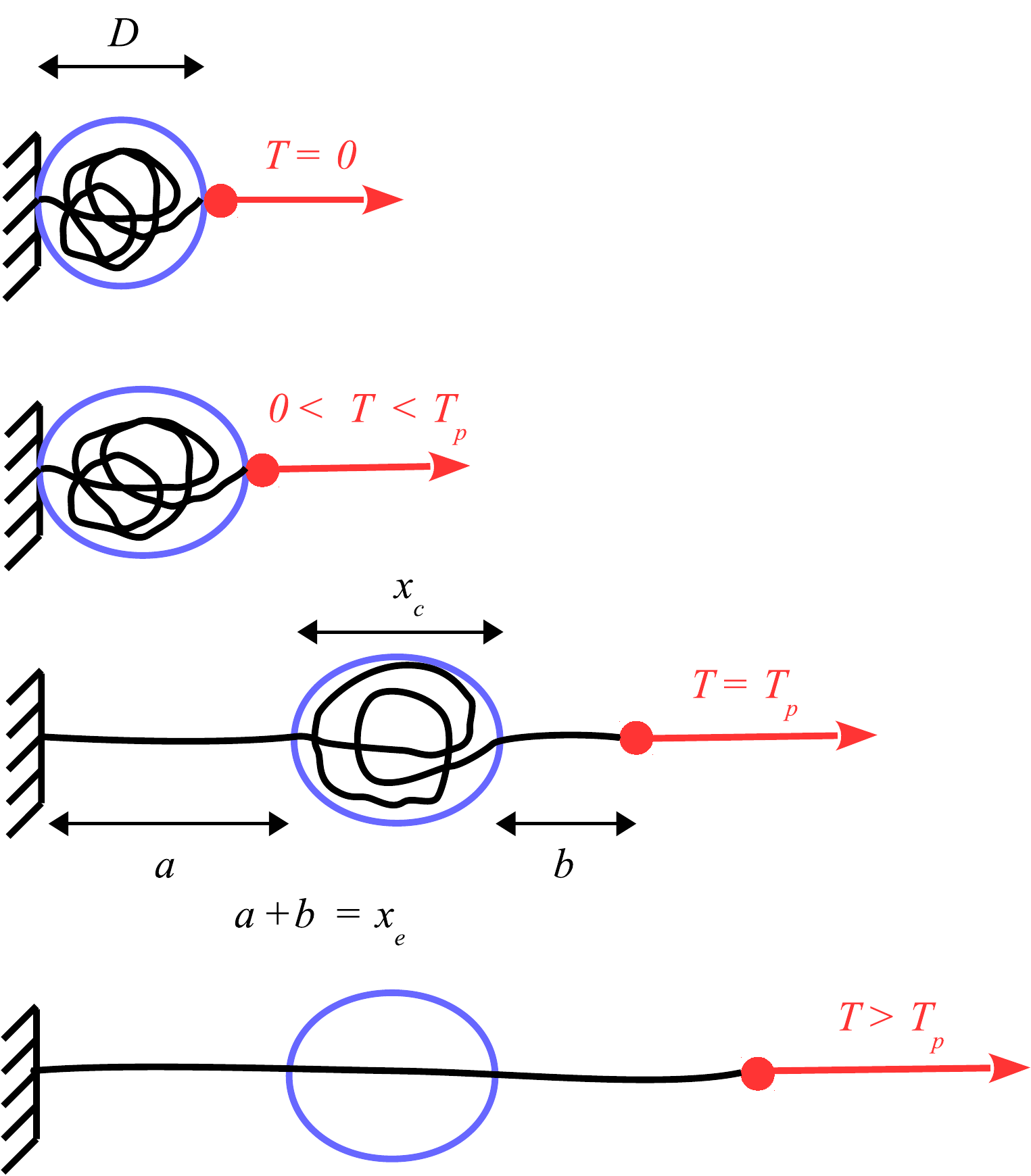}
\caption{\textbf{Two phase model.}
The windlass system comprises two phases. The extended phase consists of the fiber that lies outside the drop(s). The coiled phase consists of the the fiber that is coiled inside the drop(s). Each phase has its own extension, and there is a fraction $\rho$ (resp. $1-\rho$) of the complete fiber is the extended (resp. coiled) phase.
}
\label{fig:modele}
\end{center}
\end{figure}

The extended phase has extension $x_e$. It is modelled as a spring with stiffness $k_e = \pi \, h^2 \, E/\ell$ and rest length $\ell$, where $E$ is the Young's modulus of the fiber and $h$ the radius of its circular cross-section. 
The tension-extension relation of the extended phase is then
\begin{equation}
T=\pi \, h^2 \, E (x_e/\ell - 1)
\end{equation}
As explained in the main text, $\epsilon_0$ is the energy associated with the transformation of a unit length of the coiled phase into a unit length of the extended phase:
\begin{equation}
\epsilon_0 =2 \pi \, h \, \gamma \cos \theta - 1/2 \, \pi E \, h^4/D^2
\end{equation}
where $\theta$ is the wetting angle of the liquid on the fiber.
The total energy of the extended phase is then
\begin{equation}
{\cal E}_e = 1/2 \, k_e (x_e - \ell)^2 +\epsilon_0 \, \ell
\end{equation}
When a fraction $\rho$ (respectively $1-\rho$) of the system is in the extended (resp. coiled) phase, the global energy ${\cal E}(x_e,x_c,\rho)$ is given by:
\begin{equation}
{\cal E} = (1-\rho) \, {\cal E}_c(x_c) + \rho \, {\cal E}_e(x_e)
\end{equation}
where $\ell$ is the natural length of the fiber.

We minimize this energy under the fixed-end constraint of given global extension $x$:
\begin{equation}
x = (1-\rho) \, x_c + \rho \, x_e
\label{equa:constraint-X}
\end{equation}
Consequently we work with the function ${\cal L}={\cal E} - \lambda \, [ (1-\rho) x_c + \rho x_e]$ and solve for 
\begin{equation}
(\partial {\cal L}/\partial x_c, \partial {\cal L}/\partial x_e, \partial {\cal L}/\partial \rho)=0
\end{equation}
The Lagrange multiplier $\lambda$ is easily identified with the external force $T$ needed to enforce the constraint (\ref{equa:constraint-X}).

A first solution is $\rho=0$, $T=\lambda = k_c \, (x - D)$, and $x_c=x$; the system is entirely in the coiled phase.

A second solution is $\rho=1$, $T = \lambda =k_e \, (x - \ell)$, and $x_e=x$; the system is entirely in the extended phase.

A third solution is $0<\rho<1$, that is the system is in a mixture of phases. The tension is given by $T = {\cal E}_c'(x_c) = {\cal E}_e'(x_e)  = ({\cal E}_e - {\cal E}_c)/(x_e-x_c)$ yielding
\begin{equation}
T = \lambda = \epsilon_0 \, \left( 1 + \frac{\lambda^2}{2 \,  \epsilon_0 \, k_c} - \frac{\lambda^2}{2 \, \epsilon_0\,  k_e} \right) \Big/ \left( 1-\frac{D}{\ell} \right)
\end{equation}
which, as $D \ll L$ and $\epsilon_0 \ll k_e \, , k_c$, simplifies to $T \simeq \epsilon_0$.
\begin{acknowledgments}
The present work was supported by ANR grant  ANR-09-JCJC-0022-01, `La Ville de Paris - Programme \'Emergence', Royal Society International Exchanges Scheme 2013/R1 grant IE130506, and the PEPS PTI program from CNRS.
We thank Régis Wunenburger for discussions and experimental advices on thread visualization within droplets and Christine Rollard for advices in spider housing.
\end{acknowledgments}


\begin{thebibliography}{10}

\bibitem{Foelix2010}
Rainer Foelix.
\newblock {\em Biology of spiders}.
\newblock Oxford University Press, 2010.

\bibitem{Becker2003}
Nathan Becker, Emin Oroudjev, Stephanie Mutz, Jason~P. Cleveland, Paul~K.
  Hansma, Cheryl~Y. Hayashi, Dmitrii~E. Makarov, and Helen~G. Hansma.
\newblock Molecular nanosprings in spider capture-silk threads.
\newblock {\em Nat Mater}, 2(4):278--283, 2003.

\bibitem{Blackledge2005}
Todd~A. Blackledge, Adam~P. Summers, and Cheryl~Y. Hayashi.
\newblock Gumfooted lines in black widow cobwebs and the mechanical properties
  of spider capture silk.
\newblock {\em Zoology}, 108(1):41--46, 3 2005.

\bibitem{Vollrath1989}
Fritz Vollrath and Donald~T. Edmonds.
\newblock Modulation of the mechanical properties of spider silk by coating
  with water.
\newblock {\em Nature}, 340:305--307, 1989.

\bibitem{Schneider1995}
P.~Schneider.
\newblock Elastic properties of the viscid silk of orb weaving spiders
  (araneidae).
\newblock {\em Naturwissenschaften}, 82(3):144--145, 1995.

\bibitem{Vollrath1995}
F.~Vollrath and D.~Edmonds.
\newblock Elastic properties of spider's capture silk.
\newblock {\em Naturwissenschaften}, 82(8):379--380, 1995.

\bibitem{Peters1995}
H.M. Peters.
\newblock Ultrastructure of orb spiders' gluey capture threads.
\newblock {\em Naturwissenschaften}, 82(8):380--382, 1995.

\bibitem{Adam1937}
N.~K. Adam.
\newblock Detergent action and its relation to wetting and emulsification.
\newblock {\em Journal of the Society of Dyers and Colourists}, 53(4):121--129,
  1937.

\bibitem{Quere1999}
David Qu{\'e}r{\'e}.
\newblock Fluid coating on a fiber.
\newblock {\em Annual Review of Fluid Mechanics}, 31(1):347--384, 2011/01/31
  1999.

\bibitem{Carroll1989}
Brendan~Joseph Carroll.
\newblock Droplet formation and contact angles of liquids on mammalian hair
  fibres.
\newblock {\em Journal of the Chemical Society, Faraday Transactions 1:
  Physical Chemistry in Condensed Phases}, 85(11):3853--3860, 1989.

\bibitem{Vollrath1991}
F.~Vollrath and E.K. Tillinghast.
\newblock Glycoprotein glue beneath a spider web's aqueous coat.
\newblock {\em Naturwissenschaften}, 78(12):557--559, 1991.

\bibitem{Edmonds1992}
Donald~T. Edmonds and Fritz Vollrath.
\newblock The contribution of atmospheric water vapour to the formation and
  efficiency of a spider's capture web.
\newblock {\em Proceedings of the Royal Society of London. Series B: Biological
  Sciences}, 248(1322):145--148, 1992.

\bibitem{Gennes2003}
P.G. de~Gennes, F.~Brochard-Wyart, and D.~Qu\'er\'e.
\newblock {\em Capillarity and Wetting Phenomena: Drops, Bubbles, Pearls,
  Waves}.
\newblock Springer, 2003.

\bibitem{Lorenceau2006}
Elise Lorenceau, Tim Senden, and David Qu{\'e}r{\'e}.
\newblock Wetting of fibers.
\newblock In R.G. Weiss and P.~Terech, editors, {\em Molecular Gels. Materials
  with Self-Assembled Fibrillar Networks}, pages 223--237. Springer, 2006.

\bibitem{Marko2003}
John~F. Marko and Simona Cocco.
\newblock The micromechanics of {DNA}.
\newblock {\em Physics World}, pages 37--41, 2003.

\bibitem{Audoly2010}
Basile Audoly and Yves Pomeau.
\newblock {\em Elasticity and Geometry: From hair curls to the non-linear
  response of shells}.
\newblock Oxford University Press, 2010.

\bibitem{Antkowiak2011}
A.~Antkowiak, B.~Audoly, C.~Josserand, S.~Neukirch, and M.~Rivetti.
\newblock Instant fabrication and selection of folded structures using drop
  impact.
\newblock {\em Proc. Natl Acad. Sci. U.S.A.}, 108(26):10400--10404, 2011.

\bibitem{Gordon1978}
J.~E. Gordon.
\newblock {\em Structures or why things don't fall down}.
\newblock Penguin Books, 1978.

\bibitem{Leforestier2009}
Am{\'e}lie Leforestier and Fran{\c c}oise Livolant.
\newblock Structure of toroidal dna collapsed inside the phage capsid.
\newblock {\em Proceedings of the National Academy of Sciences}, 2009.

\bibitem{Ball2009}
Philip Ball.
\newblock {\em Shapes - Nature's Patterns: A Tapestry in Three Parts}.
\newblock Oxford University Press, 2009.

\bibitem{Lazarus2012}
A.~Lazarus, H.~C.~B. Florijn, and P.~M. Reis.
\newblock Geometry-induced rigidity in nonspherical pressurized elastic shells.
\newblock {\em Phys. Rev. Lett.}, 109:144301, Oct 2012.

\bibitem{Krieger2012}
K~Krieger.
\newblock Buckling down.
\newblock {\em Nature}, 488(7410):146--147, 2012.

\bibitem{Jacob1977}
F~Jacob.
\newblock Evolution and tinkering.
\newblock {\em Science}, 196(4295):1161--1166, 1977.

\end{thebibliography}
\end{document}